\documentclass[aps,prc,twocolumn, superscriptaddress,showkeys,nofootinbib]{revtex4-1}  

\usepackage[utf8]{inputenc}

\usepackage{graphicx,color}
\usepackage{amsmath,amssymb,amsfonts}
\usepackage{hyperref}

\usepackage{xspace}	

\begin{document}
\title{
Search for baryon junctions in e+A collisions at the Electron Ion Collider
}
\medskip
\author{Niseem~Magdy} 
\email{niseemm@gmail.com}
\affiliation{Department of Chemistry, the State University of New York, Stony Brook, New York 11794, USA}
\affiliation{Center for Frontiers in Nuclear Science at the State University of New York, Stony Brook, NY 11794, USA}
\affiliation{Department of Physics, University of Tennessee, Knoxville, TN, 37996, USA}
\author{Abhay~Deshpande} 
\affiliation{Department of Physics, Stony Brook University, Stony Brook, NY 11794, USA}
\affiliation{Physics Department, Brookhaven National Laboratory, Upton, New York 11973, USA}
\affiliation{Center for Frontiers in Nuclear Science at the State University of New York, Stony Brook, NY 11794, USA}

\author{Roy~Lacey}
\affiliation{Department of Chemistry, the State University of New York, Stony Brook, New York 11794, USA}
\affiliation{Department of Physics, Stony Brook University, Stony Brook, NY 11794, USA}

\author{Wenliang~Li}
\affiliation{Department of Physics, Stony Brook University, Stony Brook, NY 11794, USA}
\affiliation{Center for Frontiers in Nuclear Science at the State University of New York, Stony Brook, NY 11794, USA}
\affiliation{Mississippi State University, Starkville, MS}

\author{Prithwish Tribedy}
\affiliation{Physics Department, Brookhaven National Laboratory, Upton, New York 11973, USA}
\author{Zhangbu Xu}
\affiliation{Physics Department, Kent State University, Kent, Ohio, 44242, USA}
\affiliation{Physics Department, Brookhaven National Laboratory, Upton, New York 11973, USA}
\begin{abstract}
Constituent quarks in a nucleon are the essential elements in the standard ``quark model" associated with the electric charge, spin, mass, and baryon number of a nucleon. Quantum Chromodynamics (QCD) describes nucleon as a composite object containing current quarks (valence quarks and sea (anti-)quarks) and gluons. These subatomic elements and their interactions are known to contribute in complex ways to the overall nucleon spin and mass. In the early development of QCD theory in the 1970s, an alternative hypothesis postulated that the baryon number might manifest itself through a non-perturbative configuration of gluon fields forming a Y-shaped topology known as the gluon junction. In this work, we propose to test such hypothesis by measuring (i) the Regge intercept of the net-baryon distributions for $e$+($p$)Au collisions, (ii) baryon and charge transport in the isobaric ratio between $e$+Ru and $e$+Zr collisions, and (iii) target flavor dependence of proton and antiproton yields at large rapidity,  transported from the hydrogen and deuterium targets in $e+p$(d) collisions. Our study indicates that these measurements at the EIC can help determine what carries the baryon number.
\end{abstract}
\maketitle

\section{INTRODUCTION} \label{sec:2}
\begin{figure}[!h]
 \centering{
 \includegraphics[width=0.99\linewidth,angle=0]{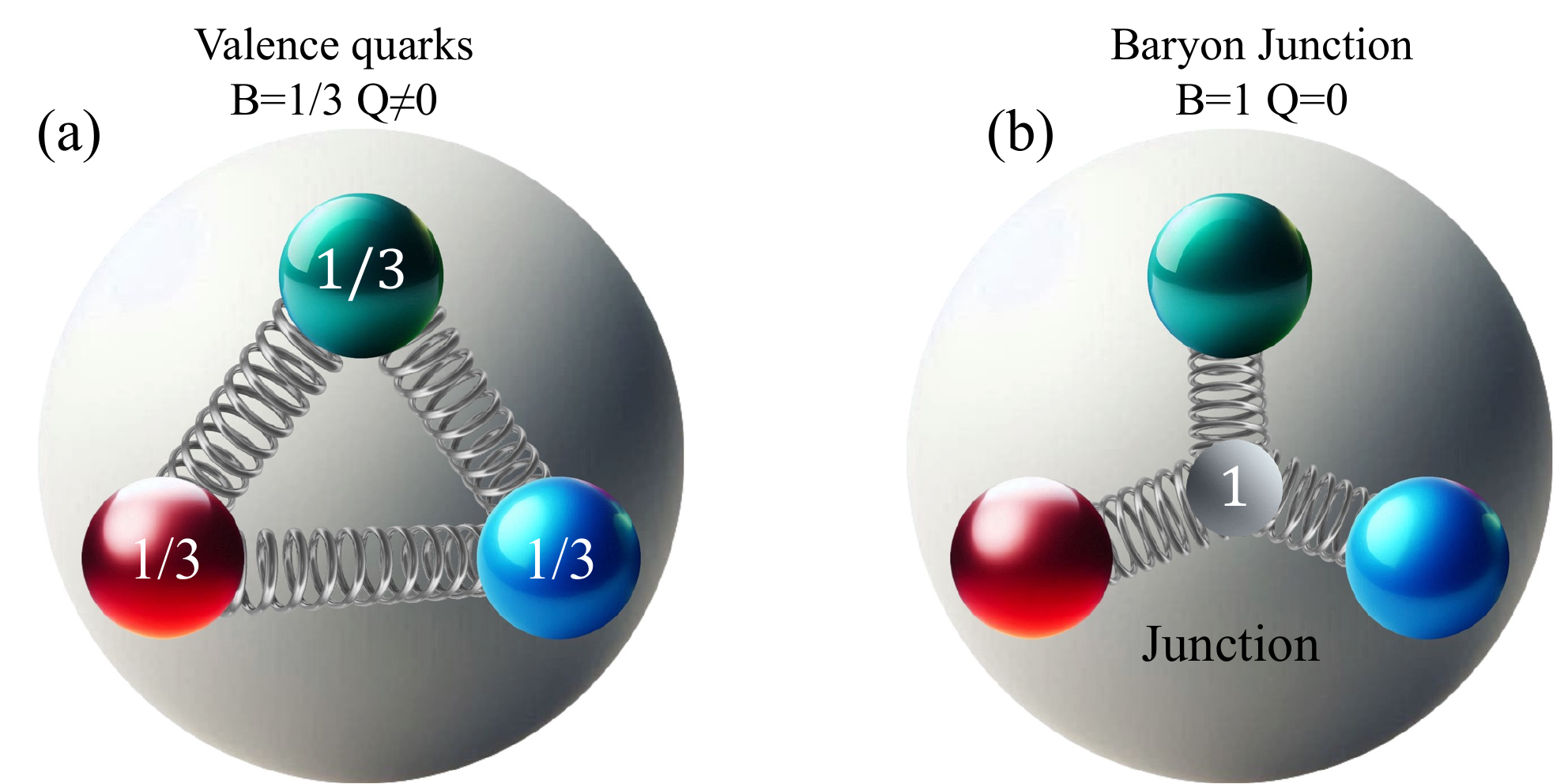}
\vskip -0.26cm
 \caption{
Schematic illustration of two types of baryon number carriers, panel (a) valence quarks and panel (b) baryon junction.
 \label{fig:00}
 }
 \vskip -0.46cm
 }
 \end{figure}

The conservation of the baryon number is a fundamental principle in physics, playing a critical role in the stability of protons, the lightest of the baryons. According to the Standard Model, quarks are the building blocks of hadrons and interact via the strong force. They are believed to each carry a baryon number of $\pm1$/$3$, summing to a total baryon number of one in the baryons (see Fig.~\ref{fig:00} (a)) and zero for mesons. Typically, these quarks are believed to interact via gluons and depicted to be in a triangular configuration ($\Delta-$ shaped)~\cite{Cornwall:1996xr}). However, an alternative theory proposed in the early 1970s suggests that the valence quarks are connected in a $Y$-shaped structure known as gluon-junction or baryon-junction~\cite{Artru:1974zn, Rossi:1977cy, Kharzeev:1996sq} (see Fig.~\ref{fig:00} (b)). This junction is assumed to be the gauge-invariant structure that carries the baryon number~\cite{Suganuma:2004zx, Takahashi:2000te}. Despite its theoretical appeal and studies in Lattice QCD, neither scenario has been conclusively verified experimentally due to the indistinguishable nature of quark and junction configurations in most physical processes.

In the context of ultra-relativistic nucleus-nucleus collisions, a substantial puzzle has been the observed baryon asymmetry at mid-rapidity, a phenomenon inconsistent with the expected transport behavior of valence quarks~\cite{Gyulassy:1997mz, NA49:1998gaz, BRAHMS:2009wlg, STAR:2008med, STAR:2017sal, ALICE:2010hjm}. The expected behavior follows from the fact that valence quarks, carrying a large fraction of the baryon momentum and therefore having shorter interaction times, should end up in the fragmentation regions far from mid-rapidity~\cite{Kharzeev:1996sq, Cebra:2022avc}. Nevertheless, the baryon junction model offers a compelling explanation for this discrepancy. Baryon junctions, composed of low-momentum gluons, can interact with the soft parton field of the projectile and become stopped at mid-rapidity.
This mechanism leads to the production of baryons with low transverse momentum and different quark content than the colliding baryons and accompanying pions, leading to high multiplicity events.
In addition, the exponential drop in cross-section with rapidity loss, characteristic of junction-stopping processes, provides a distinctive signature that supports the junction model over the valence quark model in explaining baryon transport in high-energy collisions.
 For example, in a semi-inclusive photon-induced process off a proton such as $\gamma^*+p\rightarrow B+X$, the rapidity distribution of the produced net-baryon is expected to follow 
 \begin{equation}\label{Eq:1}
   \left. \dfrac{dN}{dy} \right|_{Net-B} \sim \exp(-\alpha_J(y-Y_{\rm beam})),
\end{equation}
where $\alpha_J$ is determined by the Regge intercept of the baryon junction~\cite{Kharzeev:1996sq}. The value of $\alpha_J$ can be -0.5, 0.0, and 0.5, depending on 2,1,0  valence quarks from the colliding proton $p$ accompany the junction to form the baryon $B$ at rapidity $y$, along with the production of associated mesons $X$. These characteristic rapidity slopes of the produced baryon $B$ can serve as signatures of the baryon junction-stopping scenario~\cite{Frenklakh:2023pwy}. The valence quark picture gives rise to a much larger rapidity slope ($\alpha \sim 2.5$) in the photon-induced process as studied using Lund-based Monte-Carlo models in Ref.~\cite{Lewis:2022arg}. 

Even before the new transport mechanism of the baryon mediated by junctions was proposed~\cite{Kharzeev:1996sq}, measurements reported by the EMC Collaboration on proton transport in $\mu+p$ vs $\mu+d$~\cite{EMC_ZPC_1987} generated significant theoretical attention. This was primarily because the EMC data revealed deviations from existing models of baryon transport based on quarks and diquarks~\cite{sjostrand1986lund}. 
The data model comparisons~\cite{EMC_ZPC_1987} indicated that although the Lund model (JETSET62)~\cite{sjostrand1986lund} can qualitatively describe the dependence of proton and antiproton multiplicities on the effective mass of the hadronic final state ($W$), it failed to accurately depict their target flavor independence.
Specifically, the Lund Model~\cite{Andersson:1983ia} predicted a dependence of proton yield on the target, with a higher proton yield for the proton target compared to the deuteron target. This observation is intriguing, and detailed differential investigations of these measurements at the future EIC will be essential.

Recently, several studies have posed the question: ``Can the correlation of stopped baryons and charges be used to differentiate the two scenarios?"~\cite{Lewis:2022arg, Sinha:2023jas}. Answering this question requires an understanding of the nature of the correlations between the stopped baryons and charges. In the baryon junction scenario, the stopped charges are not correlated with the atomic number ($Z$), and the stopped baryons are expected to scale with the mass number ($A$). Conversely, in the valence quark-stopping scenario, the stopped charges are anticipated to be proportional to $Z$ and the stopped baryons to $A$ of the colliding nuclei. Consequently, it has been suggested that a relationship between the stopped baryons and charges for a particular isobaric collision could be used to test this correlation:
\begin{equation}\label{Eq:2}
    R(Isobar) = (B \times \Delta Z/A)/ \Delta Q,
\end{equation}
where $B$ and $Q$ are the average net-baryon and net-charge numbers, respectively, and $\Delta$ represents the difference between the two isobars.  Thus, $R(Isobar)$ being unity (or less than unity) can be considered a signature of the valence quark-stopping scenario, whereas $R(Isobar)$ being larger than unity is possibly a signature of the baryon junction-stopping scenario~\cite{Lv:2023fxk}.

In the 2030s, the Electron-Ion Collider (EIC) is set to advance nuclear physics research by supplying nuclear deep inelastic scattering with collider kinematics~\cite{AbdulKhalek:2021gbh}.  In the context of e+A collisions at the EIC, the reaction mechanism can be categorized into three distinct stages~\cite{Chang:2022hkt}: (i) hard scattering processes~\cite{Piller:1995kh}, (ii) intra-nuclear cascade (INC) processes~\cite{Cugnon:1982qw}, and (iii) the breakup of excited nuclear remnants~\cite{Weisskopf:1937zz}. Identifying kinematic regions dominated by each stage is crucial for refining nuclear excitation models and deepening our understanding of the target nucleus~\cite{Mathews:1982zz, Magdy:2024thf}.
Ref.~\cite{Cebra:2022avc} identifies backward ($u$-channel) production at the EIC as a promising approach for exploring baryon number transport mechanisms, potentially connected to baryon junction dynamics. Our work addresses the three suggested measurements within the EIC framework, utilizing the Benchmark eA Generator for Leptoproduction (BeAGLE) model.

The article is organized as follows. Section~\ref{sec:2} introduces the BeAGLE model and the data sets created. The results for the net-baryon distributions (Eq.~\ref{Eq:1}) and the isobaric ratios (Eq.~\ref{Eq:2}) will be presented in Section~\ref{sec:3}. Finally, a summary is provided in Section~\ref{sec:4}.

\section{MODEL AND METHODOLOGY}\label{sec:2}
\begin{figure}[!h]
 \centering{
 \includegraphics[width=0.99\linewidth,angle=0]{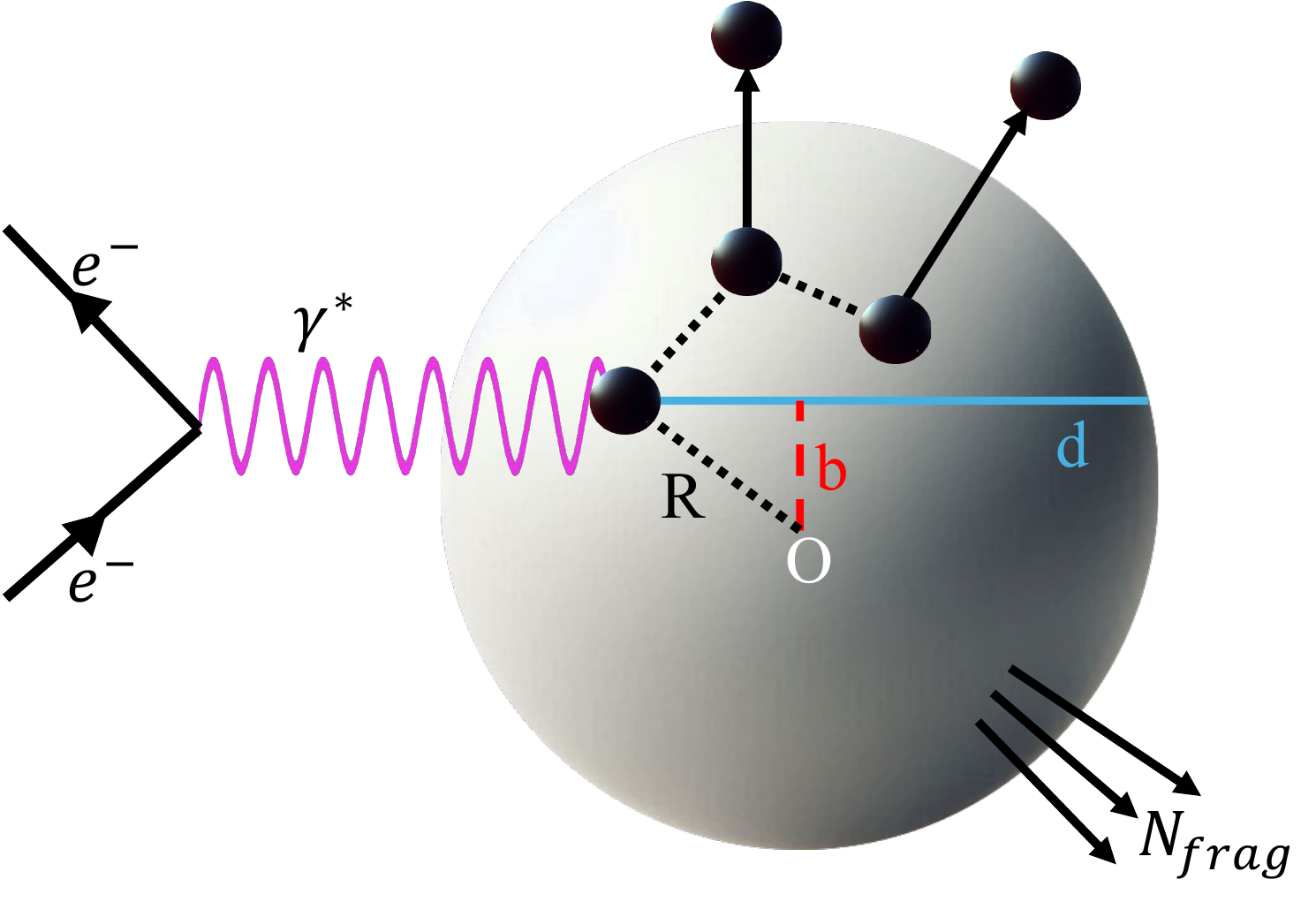}\vskip -0.26cm
 \includegraphics[width=0.99\linewidth,angle=0]{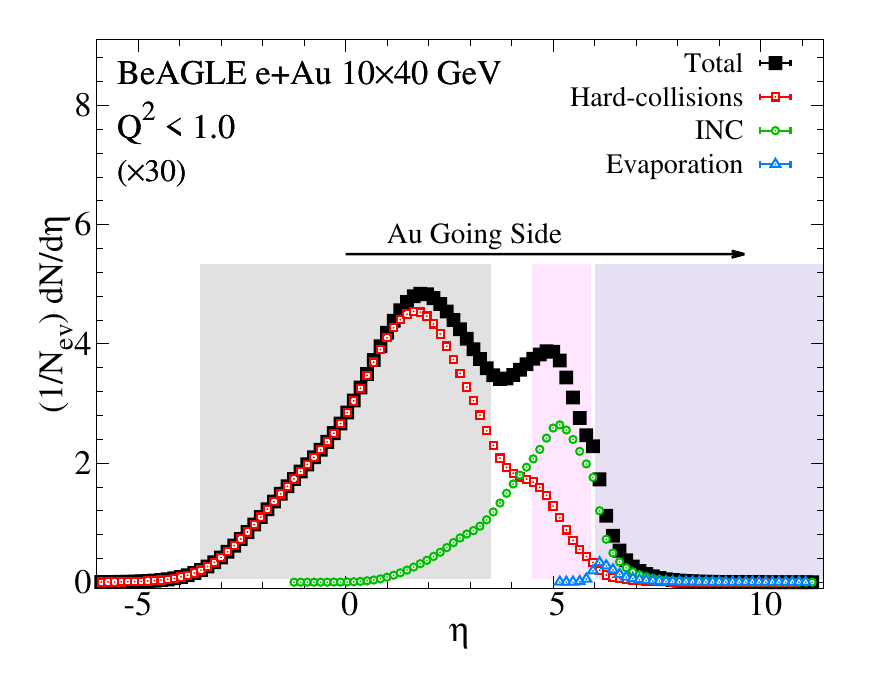}
\vskip -0.26cm
 \caption{
The top panel gives a schematic representation of the quantities describing the collision geometry, including \( b \), the impact parameter; \( R \), the spatial displacement from the interaction point to the center of the nucleus; and \( d \), the projected virtual photon travel distance from the interaction point to the edge of the nuclear medium. The lower panel displays the $dN/d\eta$ distribution of the $e$+$Au$ collisions at 10$\times$40 (GeV) from the BeAGLE model. The shaded areas represent the three $\eta$ acceptance of the three mean systems expected by the ePIC experiment at the EIC~\cite{AbdulKhalek:2021gbh, ePIC} as (i) $-3.5 < \eta < 3.5$, (ii) $4.5< \eta < 5.9$ and (iii) $\eta > 6.0$.
 \label{fig:1}
 }
 }
 \end{figure}

Our study employed version 1.03 of the Monte Carlo BeAGLE model, a versatile FORTRAN program designed for simulating electron-nucleus (eA) collisions (see Fig.~\ref{fig:1}). BeAGLE serves as a hybrid model, incorporating various established codes such as DPMJet~\cite{Roesler:2000he}, PYTHIA6~\cite{Sjostrand:2006za}, PyQM~\cite{Dupre:2011afa}, FLUKA~\cite{Bohlen:2014buj, Ferrari:2005zk}, and LHAPDF5~\cite{Whalley:2005nh} to characterize high-energy lepton-nuclear scattering phenomena comprehensively. The specific contributions of each model within BeAGLE are as follows:
\begin{itemize}
\item PYTHIA-6: Manages partonic interactions and subsequent fragmentation processes.
\item DPMJet: Defines the creation of hadrons and their interactions with the nucleus through an intra-nuclear cascade.
\item PyQM: Provides the geometric density distribution of nucleons within a nucleus and incorporates Salgado-Wiedemann quenching weights to represent partonic energy loss~\cite{SW:2003}.
\item FLUKA: Describes the decay of the excited nuclear remnant, including processes such as nucleon and light ion evaporation, nuclear fission, Fermi breakup of decay fragments, and photon emission de-excitation.
\end{itemize}
Additionally, BeAGLE includes features such as steering, multi-nucleon scattering (shadowing), and an enhanced description of the Fermi momentum distributions of nucleons within nuclei.

The analysis focuses on \( e + p(A) \) collisions at 10$\times$40 GeV, with approximately 0.3 billion events generated for each scenario considered. With the anticipated luminosity of the EIC, around 400 billion events are expected, allowing our analyses to be conducted with sufficient statistical power at the EIC.

\section{RESULTS}\label{sec:3}

In this section, we will discuss the three proposed observables: (i) the Regge intercept of the net-baryon distributions for e+(p)Au collisions, (ii) the isobaric ratio given by Eq.~\ref{Eq:2} for e+Ru and e+Zr collisions, and (iii) the EMC net-proton ratios.

\begin{figure}[!h]
\includegraphics[width=1.0  \linewidth, angle=-0,keepaspectratio=true,clip=true]{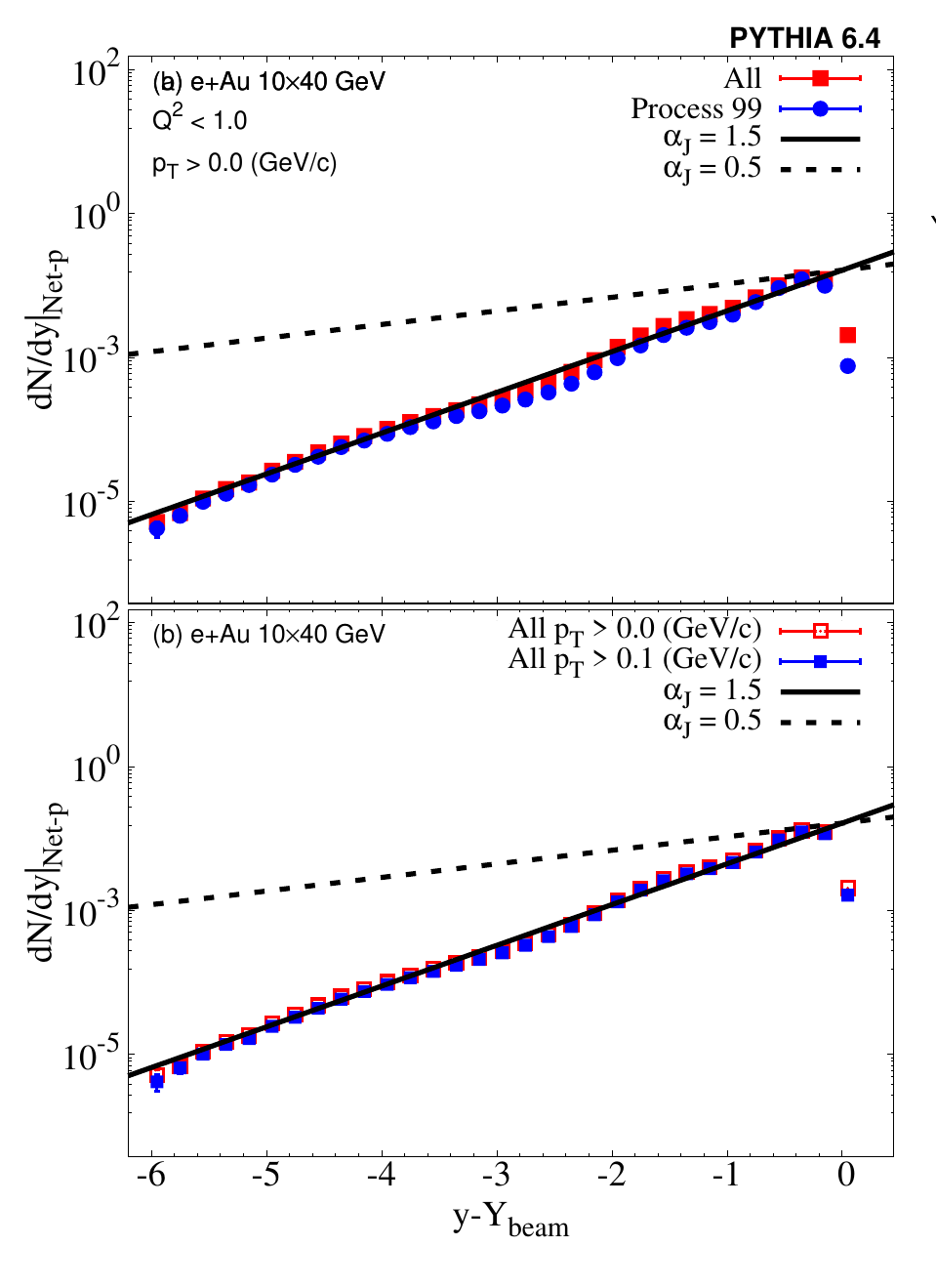}
\vskip -0.4cm
\caption{
Net-baryon rapidity distributions of e+p collisions at 10$\times$40~GeV from the PYTHIA model for all processes included and process 99 panel (a), and different $p_{T}$ cuts panel (b). The solid line represents the fit using Eq.~\ref{Eq:1}. The dashed line gives the expectation from a junction scenario~\cite{Kharzeev:1996sq}.
\label{fig:fig1}
}
\vskip -0.3cm
\end{figure}

The net-baryon distributions for e+p collisions at $10\times40$ GeV from the PYTHIA model are illustrated in Fig.~\ref{fig:fig1}. Our findings suggest that the dominant contributor to the net-baryon distribution is the PYTHIA process of LO DIS (Process 99). The results exhibit a linear relationship between the net-baryon and $y-Y_{beam}$ with a slope of $\alpha_{J} = 1.5$. It is worth noting that the net-baryon distribution from either the quark model scenario or the string junction scenario is anticipated to differ only in the slope $\alpha_{J}$. Therefore, the extracted slope of $\alpha_{J} = 1.5$ exceeds the theoretical expectation~\cite{Kharzeev:1996sq} for the baryon junction, where $\alpha_{J} =0.5$ (dashed line), and is inline with the absence of baryon junction physics in the PYTHIA model utilized in this study.
Additionally, we present the net-baryon distributions for two \( p_{T} \) cuts. Our results show minimal differences, if any, between the no-\( p_{T} \) cut and \( p_{T} < 0.1 \, \text{GeV}/c \). It is important to emphasize that hadron yields can be measured with an extrapolation of \( p_{T} \) to zero rather than being limited to the detector’s \( p_{T} \) acceptance when compared to thermal models~\cite{STAR:2024lvy, STAR:2017sal}. While uncertainties are inherent in such extrapolations, we should present a physical quantity independent of specific detector acceptance and configuration to compare to theory and other measurements.

\begin{figure}[!h] 
\includegraphics[width=1.0  \linewidth, angle=-0,keepaspectratio=true,clip=true]{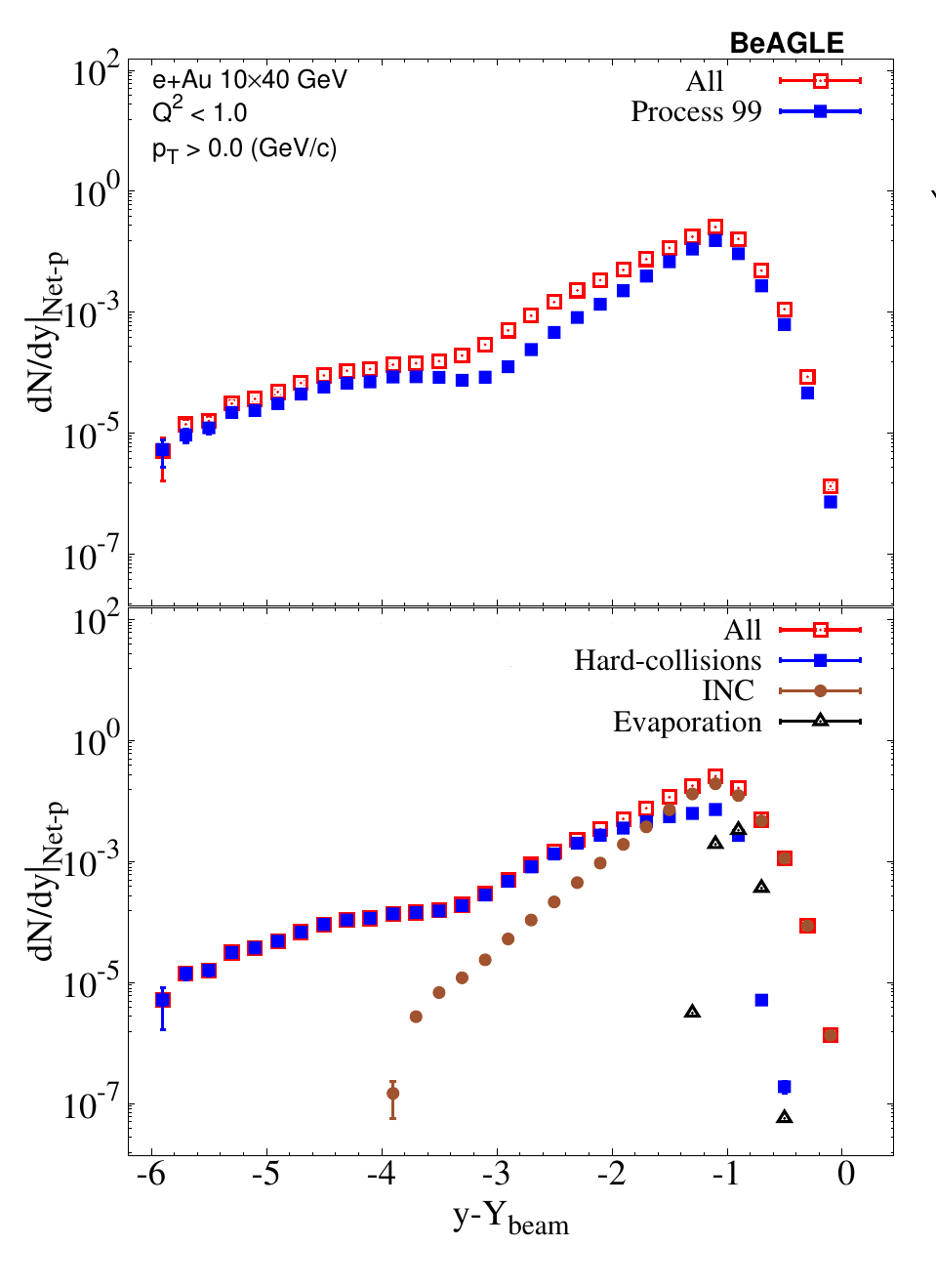}
\vskip -0.4cm
\caption{
Net-baryon rapidity distributions of e+Au collisions at 10$\times$40~GeV from the BeAGLE model for PYTHIA processes panel (a) and BeAGLE processes panel (b).
}\label{fig:fig2}
\vskip -0.3cm
\end{figure}
In contrast to the straightforward depiction presented in the e+p collisions~\ref{fig:fig1}, the measurements of net-baryon distribution in e+A collisions are anticipated to be more intricate. This anticipation stems from our understanding that the reaction mechanism in e+A collisions at the EIC encompasses various processes, including (i) hard scattering~\cite{Piller:1995kh}, (ii) IntraNuclear Cascade~\cite{Cugnon:1982qw}, and (iii) breakup of excited nuclear remnants~\cite{Weisskopf:1937zz} (see lower pane of Fig.~\ref{fig:1}). Therefore, expanding our examination of the net-baryon distribution to e+A collisions is crucial to understand its role in our baryon stooping study.
Figure~\ref{fig:fig2} displays the net-baryon distributions for e+Au collisions at $10\times40$ GeV from the BeAGLE model for multiple PYTHIA processes (a) and several BeAGLE processes (b). The net-baryon distribution results in panel (a) demonstrate varying dependencies on $y-Y_{beam}$. This variability arises from the different characteristics of particle production dependence on $y-Y_{beam}$ in the BeAGLE model given in panel (b).
Consequently, interpreting the results of the net-baryon distribution for e+Au collisions is not straightforward. Although the slope, $\alpha_{J}$, at various values of $y - Y_{beam}$ differs and exceeds theoretical expectations for the baryon junction~\cite{Kharzeev:1996sq}, further theoretical insight is needed to fully understand the baryon junction's expected behavior in e+A collisions.

%

Analyzing the associations between charge and baryon number multiplicity offers a potential avenue to distinguish between valence quarks and baryon junctions as the carriers of baryon numbers. {\color{black}Nevertheless, acquiring a precise measurement of the net charge yield poses a significant challenge due to the fact that the small net charge number is calculated based on the slight difference between the two substantial yields of positively and negatively charged particles~\cite{Lv:2023fxk}}. To address this challenge, it has been suggested to assess the difference in net charge between isobaric collisions $_{40}^{96}\mathrm{Zr}$ + $_{40}^{96}\mathrm{Zr}$ and $_{44}^{96}\mathrm{Ru}$ + $_{44}^{96}\mathrm{Ru}$~\cite{Lewis:2022arg}. In such collisions, calculating the net-charge difference ($\Delta Q = Q_{\rm{Ru}}-Q_{\rm{Zr}}$) can be achieved via double ratios by comparing positive and negative particles and then by comparing Ru+Ru collisions with Zr+Zr collisions. Therefore, one can compute the ratio $R(Isobar)$ as given in Eq.~\ref{Eq:2}. In the case of valence-quark stopping, $\Delta Q$ should be close to $B \times \Delta Z/A$ and $R(Isobar) \leq 1.0$. In contrast, in the baryon-junction stopping scenario, $\Delta Q < B \times \Delta Z/A$ and $R(Isobar) > 1.0$. {\color{black}Consequently, in this work, we provide baseline calculations of $R(Isobar)$ for e + $_{40}^{96}\mathrm{Zr}$ and e + $_{44}^{96}\mathrm{Ru}$ at the EIC kinematics.}

\begin{figure}[!h]
\includegraphics[width=1.0  \linewidth, angle=-0,keepaspectratio=true,clip=true]{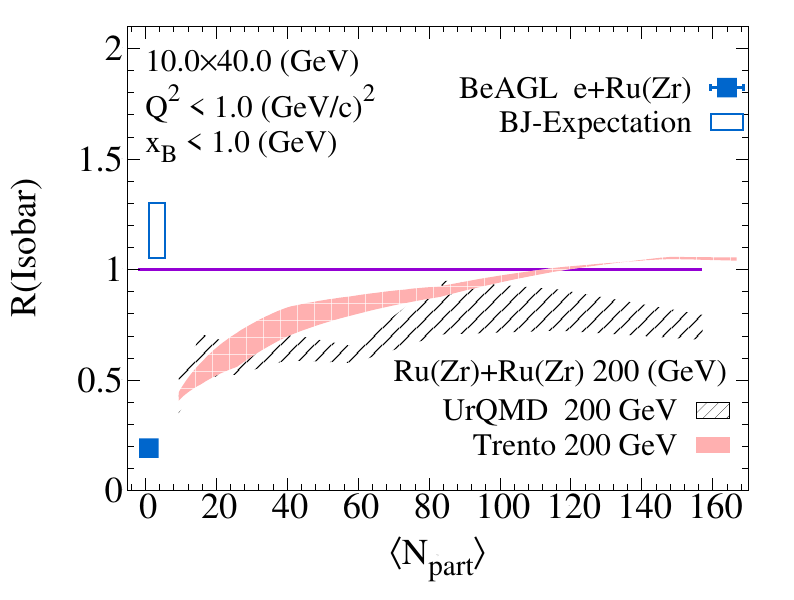}
\vskip -0.4cm
\caption{
The Isobaric ratio $(B \times \Delta Z/A)/ \Delta Q)$ as a function of $N_{part}$ in e+Ru and e+Zr collisions at 10$\times$40 GeV. Results from Ru+Ru and Zr+Zr  collisions are also shown as bands~\cite{Lv:2023fxk, Lewis:2022arg}.
}\label{fig:fig3}
\vskip -0.0cm
\end{figure}
\begin{figure}[!h] 
\includegraphics[width=1.0  \linewidth, angle=-0,keepaspectratio=true,clip=true]{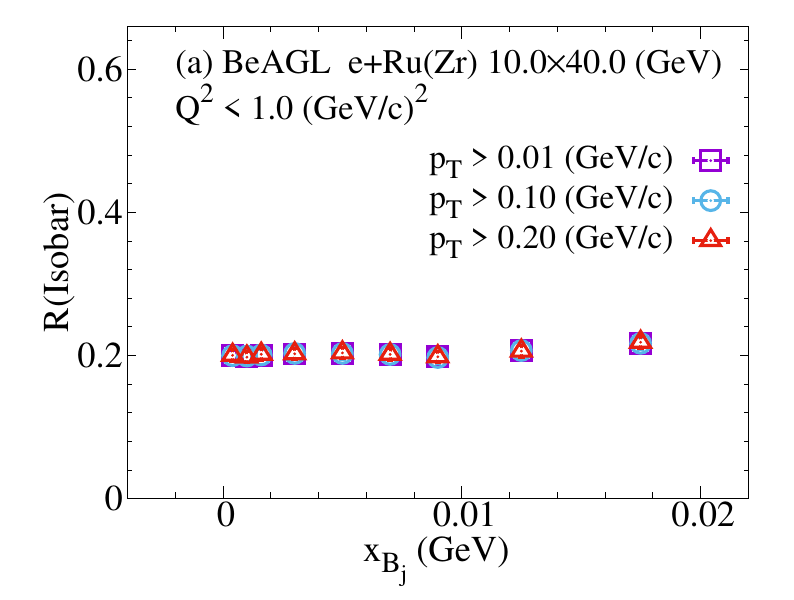}
\includegraphics[width=1.0  \linewidth, angle=-0,keepaspectratio=true,clip=true]{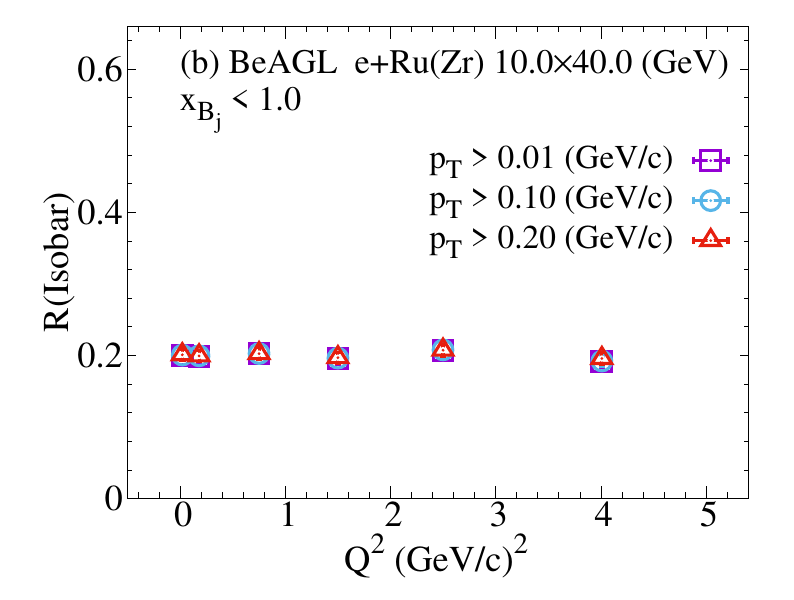}
\vskip -0.4cm
\caption{
The Isobaric ratio $(B \times \Delta Z/A)/ \Delta Q)$ as a function of $x_{B_{J}}$ panel (a) and $Q^{2}$ panel (b) in e+Ru and e+Zr collisions at 10$\times$40 GeV.
}\label{fig:fig4}
\vskip -0.3cm
\end{figure}

Figure.~\ref{fig:fig3} displays the $R(Isobar)$ vs. $N_{part}$ for (i) A+A collisions at 200 GeV and (ii) e+A collisions at 10 $\times$ 40 GeV. The results for Ru(Zr)+Ru(Zr), represented by bands from the UrQMD and Tronto models~~\cite{Lewis:2022arg, Lv:2023fxk}, show values less than unity that decrease with $N_{part}$. This behavior aligns with the scenario where valence quarks carry the baryon number. The e+Ru(Zr) results, depicted by symbols from the BeAGLE model, also show values smaller than unity and are consistent with the valence quarks carrying the baryon number.

{\color{black} 
As introduced earlier, the EIC will allow us to explore DIS in collider kinematics with wide acceptance for the Bjorken variable $x_{B_{j}}$ and the squared momentum transfer $Q^{2}$~\cite{Accardi:2012qut}.
%
Consequently, expanding our $R(Isobar)$ study into different ranges of $x_{B_{j}}$ and $Q^{2}$ is essential since the baryon junction picture is a nonperturbative QCD picture.
}
Figure.~\ref{fig:fig4} panels (a) and (b) illustrate the dependence of $R(Isobar)$ on $x_{B_{j}}$ and $Q^{2}$, respectively, for several transverse momentum selections. Our findings indicate that $R(Isobar)$ is less than unity and independent of $x_{B_{j}}$, $Q^{2}$, and $p_{T}$. These results are consistent with the valence-quarks scenario incorporated in the BeAGLE model.

\begin{figure}[!h] 
\includegraphics[width=0.99  \linewidth, angle=-0,keepaspectratio=true,clip=true]{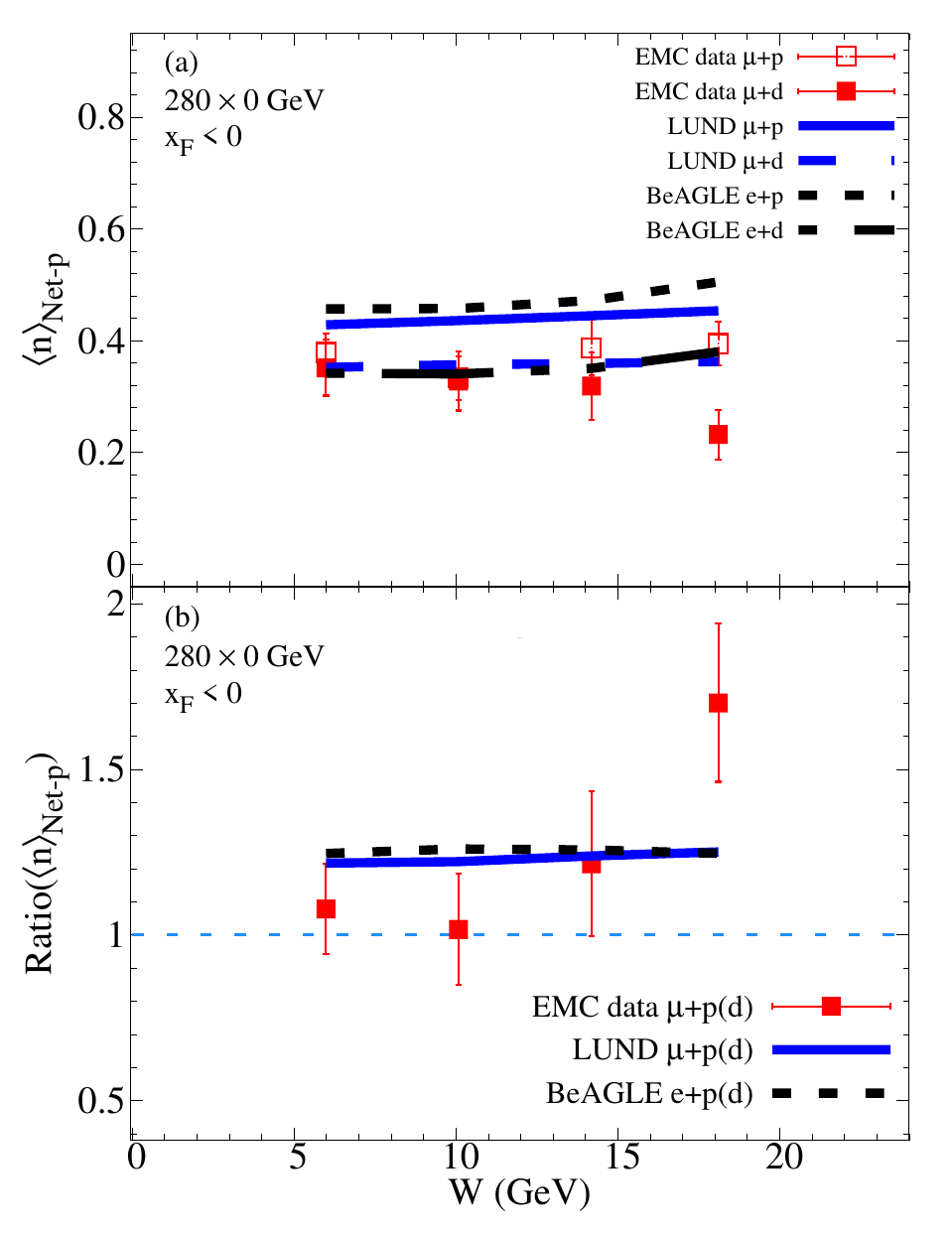}
\vskip -0.4cm
\caption{The $W$ dependence of the net-proton for $\mu$+p and $\mu$+d at 280-GeV muon on fixed targets are shown in panel (a). The ratios between $\mu$+p and $\mu$+d are presented in panel (b). Data and LUND model calculations are extracted from Ref.~\cite{EMC_ZPC_1987}.}\label{fig:EMC}
\vskip -0.3cm
\end{figure}

The EMC Collaboration previously measured proton transport in $\mu+p$ vs. $\mu+d$~\cite{EMC_ZPC_1987}. Figure~\ref{fig:EMC} presents a modified plot of these measurements, with data extracted from Ref~\cite{EMC_ZPC_1987}.
The $W$ dependence of the net-proton for $\mu$+p and $\mu$+d at 280-GeV muon on fixed targets is presented in Fig.~\ref{fig:EMC} panel (a). The EMC measurements indicated a weak dependence on the target flavor. In contrast, the LUND and BeAGLE models' calculations suggested about 20\% higher proton yield for the proton target compared to the deuteron target. Additionally, panel (b) of Fig.~\ref{fig:EMC} shows the ratio of net-proton yields between $\mu$+p and $\mu$+d from both data and simulations. The results in panel (b) indicated that while the data is target flavor independent (within the uncertainties), the simulations from the LUND and BeAGLE models revealed approximately 20\% target flavor dependence.
 
Consequently, the EMC Collaboration's measurements suggest to be inconsistent with the naive expectation that proton yield would correspond to the target flavor, meaning more protons from the proton target than the deuteron target if the baryon number is directly related to valence quarks. The EMC Collaboration stated that:`` Proton as well as antiproton production from both a hydrogen and a deuterium target are found to show the same behaviour in the $x_F$ distributions. This target independence is also true for the $W$ dependence of the average multiplicities. The multiplicities of antiprotons and forward produced protons exhibit the same significant rise with $W$. The Lund model is generally
in agreement with the data. However it predicts
a somewhat higher yield of protons from the hydrogen than from the deuterium target, an effect which is less pronounced in the data."~\cite{EMC_ZPC_1987}. Additionally, our BeAGLE model results align with the Lund model's conclusions. Therefore, more detailed measurements of the net-proton yields, with high statistics and different isobar targets at the future EIC, would help determine whether proton yields at given kinematics are correlated with the quark flavor of the target.

\section{CONCLUSION}\label{sec:4}
In summary, we investigate the potential answer to the question, ``What traces the baryon number?"~\cite{Kharzeev:1996sq} at the future EIC. We discussed (i) the Regge intercept of the net-baryon distributions in e+(p)Au collisions, (ii) the isobaric ratio constructed from the e+Ru and e+Zr collisions, and (iii) the EMC net-proton ratios at the future EIC. Our results for the extracted Regge intercept in e+(p)Au collisions indicated a value consistent with the valence-quarks scenario. The $R(Isobar)$ showed values less than unity and independent of $x_{B_{j}}$, $Q^{2}$, and momentum. Additionally, the EMC net-proton ratios challenge the expectation that proton yield would follow the target flavor in the model simulation. These simulations are consistent with the valence quarks carrying the baryon number, as included in the model we used, but appear to be inconsistent with experimental results~\cite{Lewis:2022arg, EMC_ZPC_1987} from the STAR and EMC collaborations. We conclude that conducting detailed measurements of the net-baryon distributions in e+(p)Au collisions, the isobaric ratio and the EMC net-proton ratios at the future EIC can help determine what carries the baryon number.

\section*{Acknowledgments}
We thank Rongrong Ma, Dima Kharzeev and Henry Klest for the valuable discussions, and acknowledge the attendance and discussion at the CFNS workshop titled "The 1st workshop on Baryon Dynamics from RHIC to EIC". This research is supported by the US Department of Energy, Office of Nuclear Physics (DOE NP),  under contracts DE-FG02-87ER40331.A008, DE-FG02-96ER40982, DE-FG02-89ER40531, DE-SC0012704, DE-FG02-10ER41666, and DE-AC02-98CH108.

\bibliography{ref} 

\begin{thebibliography}{37}%
\makeatletter
\providecommand \@ifxundefined [1]{%
 \@ifx{#1\undefined}
}%
\providecommand \@ifnum [1]{%
 \ifnum #1\expandafter \@firstoftwo
 \else \expandafter \@secondoftwo
 \fi
}%
\providecommand \@ifx [1]{%
 \ifx #1\expandafter \@firstoftwo
 \else \expandafter \@secondoftwo
 \fi
}%
\providecommand \natexlab [1]{#1}%
\providecommand \enquote  [1]{``#1''}%
\providecommand \bibnamefont  [1]{#1}%
\providecommand \bibfnamefont [1]{#1}%
\providecommand \citenamefont [1]{#1}%
\providecommand \href@noop [0]{\@secondoftwo}%
\providecommand \href [0]{\begingroup \@sanitize@url \@href}%
\providecommand \@href[1]{\@@startlink{#1}\@@href}%
\providecommand \@@href[1]{\endgroup#1\@@endlink}%
\providecommand \@sanitize@url [0]{\catcode `\\12\catcode `\$12\catcode
  `\&12\catcode `\#12\catcode `\^12\catcode `\_12\catcode `\%12\relax}%
\providecommand \@@startlink[1]{}%
\providecommand \@@endlink[0]{}%
\providecommand \url  [0]{\begingroup\@sanitize@url \@url }%
\providecommand \@url [1]{\endgroup\@href {#1}{\urlprefix }}%
\providecommand \urlprefix  [0]{URL }%
\providecommand \Eprint [0]{\href }%
\providecommand \doibase [0]{http://dx.doi.org/}%
\providecommand \selectlanguage [0]{\@gobble}%
\providecommand \bibinfo  [0]{\@secondoftwo}%
\providecommand \bibfield  [0]{\@secondoftwo}%
\providecommand \translation [1]{[#1]}%
\providecommand \BibitemOpen [0]{}%
\providecommand \bibitemStop [0]{}%
\providecommand \bibitemNoStop [0]{.\EOS\space}%
\providecommand \EOS [0]{\spacefactor3000\relax}%
\providecommand \BibitemShut  [1]{\csname bibitem#1\endcsname}%
\let\auto@bib@innerbib\@empty
\bibitem [{\citenamefont {Cornwall}(1996)}]{Cornwall:1996xr}%
  \BibitemOpen
  \bibfield  {author} {\bibinfo {author} {\bibfnamefont {J.~M.}\ \bibnamefont
  {Cornwall}},\ }\href {\doibase 10.1103/PhysRevD.54.6527} {\bibfield
  {journal} {\bibinfo  {journal} {Phys. Rev. D}\ }\textbf {\bibinfo {volume}
  {54}},\ \bibinfo {pages} {6527} (\bibinfo {year} {1996})},\ \Eprint
  {http://arxiv.org/abs/hep-th/9605116} {arXiv:hep-th/9605116} \BibitemShut
  {NoStop}%
\bibitem [{\citenamefont {Artru}(1975)}]{Artru:1974zn}%
  \BibitemOpen
  \bibfield  {author} {\bibinfo {author} {\bibfnamefont {X.}~\bibnamefont
  {Artru}},\ }\href {\doibase 10.1016/0550-3213(75)90019-X} {\bibfield
  {journal} {\bibinfo  {journal} {Nucl. Phys. B}\ }\textbf {\bibinfo {volume}
  {85}},\ \bibinfo {pages} {442} (\bibinfo {year} {1975})}\BibitemShut
  {NoStop}%
\bibitem [{\citenamefont {Rossi}\ and\ \citenamefont
  {Veneziano}(1977)}]{Rossi:1977cy}%
  \BibitemOpen
  \bibfield  {author} {\bibinfo {author} {\bibfnamefont {G.~C.}\ \bibnamefont
  {Rossi}}\ and\ \bibinfo {author} {\bibfnamefont {G.}~\bibnamefont
  {Veneziano}},\ }\href {\doibase 10.1016/0550-3213(77)90178-X} {\bibfield
  {journal} {\bibinfo  {journal} {Nucl. Phys. B}\ }\textbf {\bibinfo {volume}
  {123}},\ \bibinfo {pages} {507} (\bibinfo {year} {1977})}\BibitemShut
  {NoStop}%
\bibitem [{\citenamefont {Kharzeev}(1996)}]{Kharzeev:1996sq}%
  \BibitemOpen
  \bibfield  {author} {\bibinfo {author} {\bibfnamefont {D.}~\bibnamefont
  {Kharzeev}},\ }\href {\doibase 10.1016/0370-2693(96)00435-2} {\bibfield
  {journal} {\bibinfo  {journal} {Phys. Lett. B}\ }\textbf {\bibinfo {volume}
  {378}},\ \bibinfo {pages} {238} (\bibinfo {year} {1996})},\ \Eprint
  {http://arxiv.org/abs/nucl-th/9602027} {arXiv:nucl-th/9602027} \BibitemShut
  {NoStop}%
\bibitem [{\citenamefont {Suganuma}\ \emph {et~al.}(2005)\citenamefont
  {Suganuma}, \citenamefont {Takahashi}, \citenamefont {Okiharu},\ and\
  \citenamefont {Ichie}}]{Suganuma:2004zx}%
  \BibitemOpen
  \bibfield  {author} {\bibinfo {author} {\bibfnamefont {H.}~\bibnamefont
  {Suganuma}}, \bibinfo {author} {\bibfnamefont {T.~T.}\ \bibnamefont
  {Takahashi}}, \bibinfo {author} {\bibfnamefont {F.}~\bibnamefont {Okiharu}},
  \ and\ \bibinfo {author} {\bibfnamefont {H.}~\bibnamefont {Ichie}},\ }\href
  {\doibase 10.1063/1.1920939} {\bibfield  {journal} {\bibinfo  {journal} {AIP
  Conf. Proc.}\ }\textbf {\bibinfo {volume} {756}},\ \bibinfo {pages} {123}
  (\bibinfo {year} {2005})},\ \Eprint {http://arxiv.org/abs/hep-lat/0412026}
  {arXiv:hep-lat/0412026} \BibitemShut {NoStop}%
\bibitem [{\citenamefont {Takahashi}\ \emph {et~al.}(2001)\citenamefont
  {Takahashi}, \citenamefont {Matsufuru}, \citenamefont {Nemoto},\ and\
  \citenamefont {Suganuma}}]{Takahashi:2000te}%
  \BibitemOpen
  \bibfield  {author} {\bibinfo {author} {\bibfnamefont {T.~T.}\ \bibnamefont
  {Takahashi}}, \bibinfo {author} {\bibfnamefont {H.}~\bibnamefont
  {Matsufuru}}, \bibinfo {author} {\bibfnamefont {Y.}~\bibnamefont {Nemoto}}, \
  and\ \bibinfo {author} {\bibfnamefont {H.}~\bibnamefont {Suganuma}},\ }\href
  {\doibase 10.1103/PhysRevLett.86.18} {\bibfield  {journal} {\bibinfo
  {journal} {Phys. Rev. Lett.}\ }\textbf {\bibinfo {volume} {86}},\ \bibinfo
  {pages} {18} (\bibinfo {year} {2001})},\ \Eprint
  {http://arxiv.org/abs/hep-lat/0006005} {arXiv:hep-lat/0006005} \BibitemShut
  {NoStop}%
\bibitem [{\citenamefont {Gyulassy}\ \emph {et~al.}(1997)\citenamefont
  {Gyulassy}, \citenamefont {Topor~Pop},\ and\ \citenamefont
  {Vance}}]{Gyulassy:1997mz}%
  \BibitemOpen
  \bibfield  {author} {\bibinfo {author} {\bibfnamefont {M.}~\bibnamefont
  {Gyulassy}}, \bibinfo {author} {\bibfnamefont {V.}~\bibnamefont {Topor~Pop}},
  \ and\ \bibinfo {author} {\bibfnamefont {S.~E.}\ \bibnamefont {Vance}},\
  }\href {\doibase 10.1007/BF03053659} {\bibfield  {journal} {\bibinfo
  {journal} {Acta Phys. Hung. A}\ }\textbf {\bibinfo {volume} {5}},\ \bibinfo
  {pages} {299} (\bibinfo {year} {1997})},\ \Eprint
  {http://arxiv.org/abs/nucl-th/9706048} {arXiv:nucl-th/9706048} \BibitemShut
  {NoStop}%
\bibitem [{\citenamefont {Appelshauser}\ \emph {et~al.}(1999)\citenamefont
  {Appelshauser} \emph {et~al.}}]{NA49:1998gaz}%
  \BibitemOpen
  \bibfield  {author} {\bibinfo {author} {\bibfnamefont {H.}~\bibnamefont
  {Appelshauser}} \emph {et~al.} (\bibinfo {collaboration} {NA49}),\ }\href
  {\doibase 10.1103/PhysRevLett.82.2471} {\bibfield  {journal} {\bibinfo
  {journal} {Phys. Rev. Lett.}\ }\textbf {\bibinfo {volume} {82}},\ \bibinfo
  {pages} {2471} (\bibinfo {year} {1999})},\ \Eprint
  {http://arxiv.org/abs/nucl-ex/9810014} {arXiv:nucl-ex/9810014} \BibitemShut
  {NoStop}%
\bibitem [{\citenamefont {Arsene}\ \emph {et~al.}(2009)\citenamefont {Arsene}
  \emph {et~al.}}]{BRAHMS:2009wlg}%
  \BibitemOpen
  \bibfield  {author} {\bibinfo {author} {\bibfnamefont {I.~C.}\ \bibnamefont
  {Arsene}} \emph {et~al.} (\bibinfo {collaboration} {BRAHMS}),\ }\href
  {\doibase 10.1016/j.physletb.2009.05.049} {\bibfield  {journal} {\bibinfo
  {journal} {Phys. Lett. B}\ }\textbf {\bibinfo {volume} {677}},\ \bibinfo
  {pages} {267} (\bibinfo {year} {2009})},\ \Eprint
  {http://arxiv.org/abs/0901.0872} {arXiv:0901.0872 [nucl-ex]} \BibitemShut
  {NoStop}%
\bibitem [{\citenamefont {Abelev}\ \emph {et~al.}(2009)\citenamefont {Abelev}
  \emph {et~al.}}]{STAR:2008med}%
  \BibitemOpen
  \bibfield  {author} {\bibinfo {author} {\bibfnamefont {B.~I.}\ \bibnamefont
  {Abelev}} \emph {et~al.} (\bibinfo {collaboration} {STAR}),\ }\href {\doibase
  10.1103/PhysRevC.79.034909} {\bibfield  {journal} {\bibinfo  {journal} {Phys.
  Rev. C}\ }\textbf {\bibinfo {volume} {79}},\ \bibinfo {pages} {034909}
  (\bibinfo {year} {2009})},\ \Eprint {http://arxiv.org/abs/0808.2041}
  {arXiv:0808.2041 [nucl-ex]} \BibitemShut {NoStop}%
\bibitem [{\citenamefont {Adamczyk}\ \emph {et~al.}(2017)\citenamefont
  {Adamczyk} \emph {et~al.}}]{STAR:2017sal}%
  \BibitemOpen
  \bibfield  {author} {\bibinfo {author} {\bibfnamefont {L.}~\bibnamefont
  {Adamczyk}} \emph {et~al.} (\bibinfo {collaboration} {STAR}),\ }\href
  {\doibase 10.1103/PhysRevC.96.044904} {\bibfield  {journal} {\bibinfo
  {journal} {Phys. Rev. C}\ }\textbf {\bibinfo {volume} {96}},\ \bibinfo
  {pages} {044904} (\bibinfo {year} {2017})},\ \Eprint
  {http://arxiv.org/abs/1701.07065} {arXiv:1701.07065 [nucl-ex]} \BibitemShut
  {NoStop}%
\bibitem [{\citenamefont {Aamodt}\ \emph {et~al.}(2010)\citenamefont {Aamodt}
  \emph {et~al.}}]{ALICE:2010hjm}%
  \BibitemOpen
  \bibfield  {author} {\bibinfo {author} {\bibfnamefont {K.}~\bibnamefont
  {Aamodt}} \emph {et~al.} (\bibinfo {collaboration} {ALICE}),\ }\href
  {\doibase 10.1103/PhysRevLett.105.072002} {\bibfield  {journal} {\bibinfo
  {journal} {Phys. Rev. Lett.}\ }\textbf {\bibinfo {volume} {105}},\ \bibinfo
  {pages} {072002} (\bibinfo {year} {2010})},\ \Eprint
  {http://arxiv.org/abs/1006.5432} {arXiv:1006.5432 [hep-ex]} \BibitemShut
  {NoStop}%
\bibitem [{\citenamefont {Cebra}\ \emph {et~al.}(2022)\citenamefont {Cebra},
  \citenamefont {Sweger}, \citenamefont {Dong}, \citenamefont {Ji},\ and\
  \citenamefont {Klein}}]{Cebra:2022avc}%
  \BibitemOpen
  \bibfield  {author} {\bibinfo {author} {\bibfnamefont {D.}~\bibnamefont
  {Cebra}}, \bibinfo {author} {\bibfnamefont {Z.}~\bibnamefont {Sweger}},
  \bibinfo {author} {\bibfnamefont {X.}~\bibnamefont {Dong}}, \bibinfo {author}
  {\bibfnamefont {Y.}~\bibnamefont {Ji}}, \ and\ \bibinfo {author}
  {\bibfnamefont {S.~R.}\ \bibnamefont {Klein}},\ }\href {\doibase
  10.1103/PhysRevC.106.015204} {\bibfield  {journal} {\bibinfo  {journal}
  {Phys. Rev. C}\ }\textbf {\bibinfo {volume} {106}},\ \bibinfo {pages}
  {015204} (\bibinfo {year} {2022})},\ \Eprint
  {http://arxiv.org/abs/2204.07915} {arXiv:2204.07915 [hep-ph]} \BibitemShut
  {NoStop}%
\bibitem [{\citenamefont {Frenklakh}\ \emph {et~al.}(2024)\citenamefont
  {Frenklakh}, \citenamefont {Kharzeev},\ and\ \citenamefont
  {Li}}]{Frenklakh:2023pwy}%
  \BibitemOpen
  \bibfield  {author} {\bibinfo {author} {\bibfnamefont {D.}~\bibnamefont
  {Frenklakh}}, \bibinfo {author} {\bibfnamefont {D.~E.}\ \bibnamefont
  {Kharzeev}}, \ and\ \bibinfo {author} {\bibfnamefont {W.}~\bibnamefont
  {Li}},\ }\href {\doibase 10.1016/j.physletb.2024.138680} {\bibfield
  {journal} {\bibinfo  {journal} {Phys. Lett. B}\ }\textbf {\bibinfo {volume}
  {853}},\ \bibinfo {pages} {138680} (\bibinfo {year} {2024})},\ \Eprint
  {http://arxiv.org/abs/2312.15039} {arXiv:2312.15039 [hep-ph]} \BibitemShut
  {NoStop}%
\bibitem [{\citenamefont {Lewis}\ \emph {et~al.}(2024)\citenamefont {Lewis},
  \citenamefont {Lv}, \citenamefont {Ross}, \citenamefont {Tsang},
  \citenamefont {Brandenburg}, \citenamefont {Lin}, \citenamefont {Ma},
  \citenamefont {Tang}, \citenamefont {Tribedy},\ and\ \citenamefont
  {Xu}}]{Lewis:2022arg}%
  \BibitemOpen
  \bibfield  {author} {\bibinfo {author} {\bibfnamefont {N.}~\bibnamefont
  {Lewis}}, \bibinfo {author} {\bibfnamefont {W.}~\bibnamefont {Lv}}, \bibinfo
  {author} {\bibfnamefont {M.~A.}\ \bibnamefont {Ross}}, \bibinfo {author}
  {\bibfnamefont {C.~Y.}\ \bibnamefont {Tsang}}, \bibinfo {author}
  {\bibfnamefont {J.~D.}\ \bibnamefont {Brandenburg}}, \bibinfo {author}
  {\bibfnamefont {Z.-W.}\ \bibnamefont {Lin}}, \bibinfo {author} {\bibfnamefont
  {R.}~\bibnamefont {Ma}}, \bibinfo {author} {\bibfnamefont {Z.}~\bibnamefont
  {Tang}}, \bibinfo {author} {\bibfnamefont {P.}~\bibnamefont {Tribedy}}, \
  and\ \bibinfo {author} {\bibfnamefont {Z.}~\bibnamefont {Xu}},\ }\href
  {\doibase 10.1140/epjc/s10052-024-12834-2} {\bibfield  {journal} {\bibinfo
  {journal} {Eur. Phys. J. C}\ }\textbf {\bibinfo {volume} {84}},\ \bibinfo
  {pages} {590} (\bibinfo {year} {2024})},\ \Eprint
  {http://arxiv.org/abs/2205.05685} {arXiv:2205.05685 [hep-ph]} \BibitemShut
  {NoStop}%
\bibitem [{\citenamefont {{Arneodo}}\ \emph {et~al.}(1987)\citenamefont
  {{Arneodo}}, \citenamefont {{Arvidson}}, \citenamefont {{Aubert}},
  \citenamefont {{Badelek}}, \citenamefont {{Beaufays}} \emph
  {et~al.}}]{EMC_ZPC_1987}%
  \BibitemOpen
  \bibfield  {author} {\bibinfo {author} {\bibfnamefont {M.}~\bibnamefont
  {{Arneodo}}}, \bibinfo {author} {\bibfnamefont {A.}~\bibnamefont
  {{Arvidson}}}, \bibinfo {author} {\bibfnamefont {J.~J.}\ \bibnamefont
  {{Aubert}}}, \bibinfo {author} {\bibfnamefont {B.}~\bibnamefont {{Badelek}}},
  \bibinfo {author} {\bibfnamefont {J.}~\bibnamefont {{Beaufays}}},  \emph
  {et~al.},\ }\href {\doibase 10.1007/BF01596894} {\bibfield  {journal}
  {\bibinfo  {journal} {Zeitschrift fur Physik C Particles and Fields}\
  }\textbf {\bibinfo {volume} {35}},\ \bibinfo {pages} {433} (\bibinfo {year}
  {1987})}\BibitemShut {NoStop}%
\bibitem [{\citenamefont {Sj{\"o}strand}(1986)}]{sjostrand1986lund}%
  \BibitemOpen
  \bibfield  {author} {\bibinfo {author} {\bibfnamefont {T.}~\bibnamefont
  {Sj{\"o}strand}},\ }\href@noop {} {\bibfield  {journal} {\bibinfo  {journal}
  {Computer Physics Communications}\ }\textbf {\bibinfo {volume} {39}},\
  \bibinfo {pages} {347} (\bibinfo {year} {1986})}\BibitemShut {NoStop}%
\bibitem [{\citenamefont {Andersson}\ \emph {et~al.}(1983)\citenamefont
  {Andersson}, \citenamefont {Gustafson}, \citenamefont {Ingelman},\ and\
  \citenamefont {Sjostrand}}]{Andersson:1983ia}%
  \BibitemOpen
  \bibfield  {author} {\bibinfo {author} {\bibfnamefont {B.}~\bibnamefont
  {Andersson}}, \bibinfo {author} {\bibfnamefont {G.}~\bibnamefont
  {Gustafson}}, \bibinfo {author} {\bibfnamefont {G.}~\bibnamefont {Ingelman}},
  \ and\ \bibinfo {author} {\bibfnamefont {T.}~\bibnamefont {Sjostrand}},\
  }\href {\doibase 10.1016/0370-1573(83)90080-7} {\bibfield  {journal}
  {\bibinfo  {journal} {Phys. Rept.}\ }\textbf {\bibinfo {volume} {97}},\
  \bibinfo {pages} {31} (\bibinfo {year} {1983})}\BibitemShut {NoStop}%
\bibitem [{\citenamefont {Sinha}\ \emph {et~al.}(2023)\citenamefont {Sinha},
  \citenamefont {Bairathi}, \citenamefont {Gopal}, \citenamefont {Jena},\ and\
  \citenamefont {Kabana}}]{Sinha:2023jas}%
  \BibitemOpen
  \bibfield  {author} {\bibinfo {author} {\bibfnamefont {P.}~\bibnamefont
  {Sinha}}, \bibinfo {author} {\bibfnamefont {V.}~\bibnamefont {Bairathi}},
  \bibinfo {author} {\bibfnamefont {K.}~\bibnamefont {Gopal}}, \bibinfo
  {author} {\bibfnamefont {C.}~\bibnamefont {Jena}}, \ and\ \bibinfo {author}
  {\bibfnamefont {S.}~\bibnamefont {Kabana}},\ }\href {\doibase
  10.1103/PhysRevC.108.024911} {\bibfield  {journal} {\bibinfo  {journal}
  {Phys. Rev. C}\ }\textbf {\bibinfo {volume} {108}},\ \bibinfo {pages}
  {024911} (\bibinfo {year} {2023})},\ \Eprint
  {http://arxiv.org/abs/2305.13950} {arXiv:2305.13950 [hep-ph]} \BibitemShut
  {NoStop}%
\bibitem [{\citenamefont {Lv}\ \emph {et~al.}(2024)\citenamefont {Lv},
  \citenamefont {Li}, \citenamefont {Li}, \citenamefont {Ma}, \citenamefont
  {Tang}, \citenamefont {Tribedy}, \citenamefont {Tsang}, \citenamefont {Xu},\
  and\ \citenamefont {Zha}}]{Lv:2023fxk}%
  \BibitemOpen
  \bibfield  {author} {\bibinfo {author} {\bibfnamefont {W.}~\bibnamefont
  {Lv}}, \bibinfo {author} {\bibfnamefont {Y.}~\bibnamefont {Li}}, \bibinfo
  {author} {\bibfnamefont {Z.}~\bibnamefont {Li}}, \bibinfo {author}
  {\bibfnamefont {R.}~\bibnamefont {Ma}}, \bibinfo {author} {\bibfnamefont
  {Z.}~\bibnamefont {Tang}}, \bibinfo {author} {\bibfnamefont {P.}~\bibnamefont
  {Tribedy}}, \bibinfo {author} {\bibfnamefont {C.~Y.}\ \bibnamefont {Tsang}},
  \bibinfo {author} {\bibfnamefont {Z.}~\bibnamefont {Xu}}, \ and\ \bibinfo
  {author} {\bibfnamefont {W.}~\bibnamefont {Zha}},\ }\href {\doibase
  10.1088/1674-1137/ad243f} {\bibfield  {journal} {\bibinfo  {journal} {Chin.
  Phys. C}\ }\textbf {\bibinfo {volume} {48}},\ \bibinfo {pages} {044001}
  (\bibinfo {year} {2024})},\ \Eprint {http://arxiv.org/abs/2309.06445}
  {arXiv:2309.06445 [nucl-th]} \BibitemShut {NoStop}%
\bibitem [{\citenamefont {Abdul~Khalek}\ \emph {et~al.}(2022)\citenamefont
  {Abdul~Khalek} \emph {et~al.}}]{AbdulKhalek:2021gbh}%
  \BibitemOpen
  \bibfield  {author} {\bibinfo {author} {\bibfnamefont {R.}~\bibnamefont
  {Abdul~Khalek}} \emph {et~al.},\ }\href {\doibase
  10.1016/j.nuclphysa.2022.122447} {\bibfield  {journal} {\bibinfo  {journal}
  {Nucl. Phys. A}\ }\textbf {\bibinfo {volume} {1026}},\ \bibinfo {pages}
  {122447} (\bibinfo {year} {2022})},\ \Eprint
  {http://arxiv.org/abs/2103.05419} {arXiv:2103.05419 [physics.ins-det]}
  \BibitemShut {NoStop}%
\bibitem [{\citenamefont {Chang}\ \emph {et~al.}(2022)\citenamefont {Chang},
  \citenamefont {Aschenauer}, \citenamefont {Baker}, \citenamefont {Jentsch},
  \citenamefont {Lee}, \citenamefont {Tu}, \citenamefont {Yin},\ and\
  \citenamefont {Zheng}}]{Chang:2022hkt}%
  \BibitemOpen
  \bibfield  {author} {\bibinfo {author} {\bibfnamefont {W.}~\bibnamefont
  {Chang}}, \bibinfo {author} {\bibfnamefont {E.-C.}\ \bibnamefont
  {Aschenauer}}, \bibinfo {author} {\bibfnamefont {M.~D.}\ \bibnamefont
  {Baker}}, \bibinfo {author} {\bibfnamefont {A.}~\bibnamefont {Jentsch}},
  \bibinfo {author} {\bibfnamefont {J.-H.}\ \bibnamefont {Lee}}, \bibinfo
  {author} {\bibfnamefont {Z.}~\bibnamefont {Tu}}, \bibinfo {author}
  {\bibfnamefont {Z.}~\bibnamefont {Yin}}, \ and\ \bibinfo {author}
  {\bibfnamefont {L.}~\bibnamefont {Zheng}},\ }\href {\doibase
  10.1103/PhysRevD.106.012007} {\bibfield  {journal} {\bibinfo  {journal}
  {Phys. Rev. D}\ }\textbf {\bibinfo {volume} {106}},\ \bibinfo {pages}
  {012007} (\bibinfo {year} {2022})},\ \Eprint
  {http://arxiv.org/abs/2204.11998} {arXiv:2204.11998 [physics.comp-ph]}
  \BibitemShut {NoStop}%
\bibitem [{\citenamefont {Piller}\ \emph {et~al.}(1995)\citenamefont {Piller},
  \citenamefont {Ratzka},\ and\ \citenamefont {Weise}}]{Piller:1995kh}%
  \BibitemOpen
  \bibfield  {author} {\bibinfo {author} {\bibfnamefont {G.}~\bibnamefont
  {Piller}}, \bibinfo {author} {\bibfnamefont {W.}~\bibnamefont {Ratzka}}, \
  and\ \bibinfo {author} {\bibfnamefont {W.}~\bibnamefont {Weise}},\ }\href
  {\doibase 10.1007/BF01299761} {\bibfield  {journal} {\bibinfo  {journal} {Z.
  Phys. A}\ }\textbf {\bibinfo {volume} {352}},\ \bibinfo {pages} {427}
  (\bibinfo {year} {1995})},\ \Eprint {http://arxiv.org/abs/hep-ph/9504407}
  {arXiv:hep-ph/9504407} \BibitemShut {NoStop}%
\bibitem [{\citenamefont {Cugnon}(1982)}]{Cugnon:1982qw}%
  \BibitemOpen
  \bibfield  {author} {\bibinfo {author} {\bibfnamefont {J.}~\bibnamefont
  {Cugnon}},\ }\href {\doibase 10.1016/0375-9474(82)90200-7} {\bibfield
  {journal} {\bibinfo  {journal} {Nucl. Phys. A}\ }\textbf {\bibinfo {volume}
  {387}},\ \bibinfo {pages} {191C} (\bibinfo {year} {1982})}\BibitemShut
  {NoStop}%
\bibitem [{\citenamefont {Weisskopf}(1937)}]{Weisskopf:1937zz}%
  \BibitemOpen
  \bibfield  {author} {\bibinfo {author} {\bibfnamefont {V.}~\bibnamefont
  {Weisskopf}},\ }\href {\doibase 10.1103/PhysRev.52.295} {\bibfield  {journal}
  {\bibinfo  {journal} {Phys. Rev.}\ }\textbf {\bibinfo {volume} {52}},\
  \bibinfo {pages} {295} (\bibinfo {year} {1937})}\BibitemShut {NoStop}%
\bibitem [{\citenamefont {Mathews}\ \emph {et~al.}(1982)\citenamefont
  {Mathews}, \citenamefont {Glagola}, \citenamefont {Moyle},\ and\
  \citenamefont {Viola}}]{Mathews:1982zz}%
  \BibitemOpen
  \bibfield  {author} {\bibinfo {author} {\bibfnamefont {G.~J.}\ \bibnamefont
  {Mathews}}, \bibinfo {author} {\bibfnamefont {B.~G.}\ \bibnamefont
  {Glagola}}, \bibinfo {author} {\bibfnamefont {R.~A.}\ \bibnamefont {Moyle}},
  \ and\ \bibinfo {author} {\bibfnamefont {V.~E.}\ \bibnamefont {Viola}},\
  }\href {\doibase 10.1103/PhysRevC.25.2181} {\bibfield  {journal} {\bibinfo
  {journal} {Phys. Rev. C}\ }\textbf {\bibinfo {volume} {25}},\ \bibinfo
  {pages} {2181} (\bibinfo {year} {1982})}\BibitemShut {NoStop}%
\bibitem [{\citenamefont {Magdy}\ \emph {et~al.}(2024)\citenamefont {Magdy},
  \citenamefont {Hegazy}, \citenamefont {Rafaat}, \citenamefont {Li},
  \citenamefont {Deshpande}, \citenamefont {Abdelhady}, \citenamefont
  {Ellithi}, \citenamefont {Lacey},\ and\ \citenamefont {Tu}}]{Magdy:2024thf}%
  \BibitemOpen
  \bibfield  {author} {\bibinfo {author} {\bibfnamefont {N.}~\bibnamefont
  {Magdy}}, \bibinfo {author} {\bibfnamefont {M.}~\bibnamefont {Hegazy}},
  \bibinfo {author} {\bibfnamefont {A.}~\bibnamefont {Rafaat}}, \bibinfo
  {author} {\bibfnamefont {W.}~\bibnamefont {Li}}, \bibinfo {author}
  {\bibfnamefont {A.}~\bibnamefont {Deshpande}}, \bibinfo {author}
  {\bibfnamefont {A.~M.~H.}\ \bibnamefont {Abdelhady}}, \bibinfo {author}
  {\bibfnamefont {A.~Y.}\ \bibnamefont {Ellithi}}, \bibinfo {author}
  {\bibfnamefont {R.~A.}\ \bibnamefont {Lacey}}, \ and\ \bibinfo {author}
  {\bibfnamefont {Z.}~\bibnamefont {Tu}},\ }\href@noop {} {\  (\bibinfo {year}
  {2024})},\ \Eprint {http://arxiv.org/abs/2405.07844} {arXiv:2405.07844
  [nucl-th]} \BibitemShut {NoStop}%
\bibitem [{\citenamefont {Roesler}\ \emph {et~al.}(2000)\citenamefont
  {Roesler}, \citenamefont {Engel},\ and\ \citenamefont
  {Ranft}}]{Roesler:2000he}%
  \BibitemOpen
  \bibfield  {author} {\bibinfo {author} {\bibfnamefont {S.}~\bibnamefont
  {Roesler}}, \bibinfo {author} {\bibfnamefont {R.}~\bibnamefont {Engel}}, \
  and\ \bibinfo {author} {\bibfnamefont {J.}~\bibnamefont {Ranft}},\ }in\ \href
  {\doibase 10.1007/978-3-642-18211-2_166} {\emph {\bibinfo {booktitle}
  {{International Conference on Advanced Monte Carlo for Radiation Physics,
  Particle Transport Simulation and Applications (MC 2000)}}}}\ (\bibinfo
  {year} {2000})\ pp.\ \bibinfo {pages} {1033--1038},\ \Eprint
  {http://arxiv.org/abs/hep-ph/0012252} {arXiv:hep-ph/0012252} \BibitemShut
  {NoStop}%
\bibitem [{\citenamefont {Sjostrand}\ \emph {et~al.}(2006)\citenamefont
  {Sjostrand}, \citenamefont {Mrenna},\ and\ \citenamefont
  {Skands}}]{Sjostrand:2006za}%
  \BibitemOpen
  \bibfield  {author} {\bibinfo {author} {\bibfnamefont {T.}~\bibnamefont
  {Sjostrand}}, \bibinfo {author} {\bibfnamefont {S.}~\bibnamefont {Mrenna}}, \
  and\ \bibinfo {author} {\bibfnamefont {P.~Z.}\ \bibnamefont {Skands}},\
  }\href {\doibase 10.1088/1126-6708/2006/05/026} {\bibfield  {journal}
  {\bibinfo  {journal} {JHEP}\ }\textbf {\bibinfo {volume} {05}},\ \bibinfo
  {pages} {026} (\bibinfo {year} {2006})},\ \Eprint
  {http://arxiv.org/abs/hep-ph/0603175} {arXiv:hep-ph/0603175} \BibitemShut
  {NoStop}%
\bibitem [{\citenamefont {Dupr\'e}(2011)}]{Dupre:2011afa}%
  \BibitemOpen
  \bibfield  {author} {\bibinfo {author} {\bibfnamefont {R.}~\bibnamefont
  {Dupr\'e}},\ }\emph {\bibinfo {title} {{Quark Fragmentation and Hadron
  Formation in Nuclear Matter}}},\ \href@noop {} {Ph.D. thesis},\ \bibinfo
  {school} {Lyon, IPN} (\bibinfo {year} {2011})\BibitemShut {NoStop}%
\bibitem [{\citenamefont {B\"ohlen}\ \emph {et~al.}(2014)\citenamefont
  {B\"ohlen}, \citenamefont {Cerutti}, \citenamefont {Chin}, \citenamefont
  {Fass\`o}, \citenamefont {Ferrari}, \citenamefont {Ortega}, \citenamefont
  {Mairani}, \citenamefont {Sala}, \citenamefont {Smirnov},\ and\ \citenamefont
  {Vlachoudis}}]{Bohlen:2014buj}%
  \BibitemOpen
  \bibfield  {author} {\bibinfo {author} {\bibfnamefont {T.~T.}\ \bibnamefont
  {B\"ohlen}}, \bibinfo {author} {\bibfnamefont {F.}~\bibnamefont {Cerutti}},
  \bibinfo {author} {\bibfnamefont {M.~P.~W.}\ \bibnamefont {Chin}}, \bibinfo
  {author} {\bibfnamefont {A.}~\bibnamefont {Fass\`o}}, \bibinfo {author}
  {\bibfnamefont {A.}~\bibnamefont {Ferrari}}, \bibinfo {author} {\bibfnamefont
  {P.~G.}\ \bibnamefont {Ortega}}, \bibinfo {author} {\bibfnamefont
  {A.}~\bibnamefont {Mairani}}, \bibinfo {author} {\bibfnamefont {P.~R.}\
  \bibnamefont {Sala}}, \bibinfo {author} {\bibfnamefont {G.}~\bibnamefont
  {Smirnov}}, \ and\ \bibinfo {author} {\bibfnamefont {V.}~\bibnamefont
  {Vlachoudis}},\ }\href {\doibase 10.1016/j.nds.2014.07.049} {\bibfield
  {journal} {\bibinfo  {journal} {Nucl. Data Sheets}\ }\textbf {\bibinfo
  {volume} {120}},\ \bibinfo {pages} {211} (\bibinfo {year}
  {2014})}\BibitemShut {NoStop}%
\bibitem [{\citenamefont {Ferrari}\ \emph {et~al.}(2005)\citenamefont
  {Ferrari}, \citenamefont {Sala}, \citenamefont {Fasso},\ and\ \citenamefont
  {Ranft}}]{Ferrari:2005zk}%
  \BibitemOpen
  \bibfield  {author} {\bibinfo {author} {\bibfnamefont {A.}~\bibnamefont
  {Ferrari}}, \bibinfo {author} {\bibfnamefont {P.~R.}\ \bibnamefont {Sala}},
  \bibinfo {author} {\bibfnamefont {A.}~\bibnamefont {Fasso}}, \ and\ \bibinfo
  {author} {\bibfnamefont {J.}~\bibnamefont {Ranft}},\ }\href {\doibase
  10.2172/877507} {\  (\bibinfo {year} {2005}),\ 10.2172/877507}\BibitemShut
  {NoStop}%
\bibitem [{\citenamefont {Whalley}\ \emph {et~al.}(2005)\citenamefont
  {Whalley}, \citenamefont {Bourilkov},\ and\ \citenamefont
  {Group}}]{Whalley:2005nh}%
  \BibitemOpen
  \bibfield  {author} {\bibinfo {author} {\bibfnamefont {M.~R.}\ \bibnamefont
  {Whalley}}, \bibinfo {author} {\bibfnamefont {D.}~\bibnamefont {Bourilkov}},
  \ and\ \bibinfo {author} {\bibfnamefont {R.~C.}\ \bibnamefont {Group}},\ }in\
  \href@noop {} {\emph {\bibinfo {booktitle} {{HERA and the LHC: A Workshop on
  the Implications of HERA and LHC Physics (Startup Meeting, CERN, 26-27 March
  2004; Midterm Meeting, CERN, 11-13 October 2004)}}}}\ (\bibinfo {year}
  {2005})\ pp.\ \bibinfo {pages} {575--581},\ \Eprint
  {http://arxiv.org/abs/hep-ph/0508110} {arXiv:hep-ph/0508110} \BibitemShut
  {NoStop}%
\bibitem [{\citenamefont {Salgado}\ and\ \citenamefont
  {Wiedemann}(2003)}]{SW:2003}%
  \BibitemOpen
  \bibfield  {author} {\bibinfo {author} {\bibfnamefont {C.}~\bibnamefont
  {Salgado}}\ and\ \bibinfo {author} {\bibfnamefont {U.~A.}\ \bibnamefont
  {Wiedemann}},\ }\href {\doibase 10.1103/PhysRevD.68.014008} {\bibfield
  {journal} {\bibinfo  {journal} {Phys. Rev. D}\ }\textbf {\bibinfo {volume}
  {68}},\ \bibinfo {pages} {014008} (\bibinfo {year} {2003})},\ \Eprint
  {http://arxiv.org/abs/hep-ph/0302184} {arXiv:hep-ph/0302184} \BibitemShut
  {NoStop}%
\bibitem [{ePI()}]{ePIC}%
  \BibitemOpen
  \href@noop {} {\enquote {\bibinfo {title} {The epic collaboration},}\
  }\bibinfo {howpublished} {\url{https://www.epic-eic.org/}}\BibitemShut
  {NoStop}%
\bibitem [{STA(2024)}]{STAR:2024lvy}%
  \BibitemOpen
  \href@noop {} {\  (\bibinfo {year} {2024})},\ \Eprint
  {http://arxiv.org/abs/2408.15441} {arXiv:2408.15441 [nucl-ex]} \BibitemShut
  {NoStop}%
\bibitem [{\citenamefont {Accardi}\ \emph {et~al.}(2016)\citenamefont {Accardi}
  \emph {et~al.}}]{Accardi:2012qut}%
  \BibitemOpen
  \bibfield  {author} {\bibinfo {author} {\bibfnamefont {A.}~\bibnamefont
  {Accardi}} \emph {et~al.},\ }\href {\doibase 10.1140/epja/i2016-16268-9}
  {\bibfield  {journal} {\bibinfo  {journal} {Eur. Phys. J. A}\ }\textbf
  {\bibinfo {volume} {52}},\ \bibinfo {pages} {268} (\bibinfo {year} {2016})},\
  \Eprint {http://arxiv.org/abs/1212.1701} {arXiv:1212.1701 [nucl-ex]}
  \BibitemShut {NoStop}%
\end{thebibliography}%
\end{document}